\documentstyle[emulateapj]{article}
 
\def\etal{{\it et al.\/}}

\begin{document}
 
\slugcomment{Accepted by ApJ Letters}
 
\title{Improved Parameters and New Lensed Features for Q0957+561
        from WFPC2 Imaging
\footnote{Based on observations with the NASA/ESA {\it Hubble Space
Telescope}
obtained at the Space Telescope Science Institute, which is operated by the
Association of Universities for Research in Astronomy Inc., under NASA
contract NAS 5-26555.}}

\author{Gary Bernstein and Philippe Fischer\footnote{Hubble Fellow}}
\affil{Dept. of Astronomy, 830 Dennison Bldg.,
University of Michigan, Ann Arbor, MI 48109}
 
\author{ J. Anthony Tyson}
\affil{Bell Laboratories, Lucent Technologies,
700 Mountain Ave., Murray Hill, NJ 07974}
 
\author{George Rhee}
\affil{Dept. of Physics, University of Nevada Las Vegas, Las Vegas, NV 89195}
 
\begin{abstract}
New HST WFPC2 observations of the lensed double QSO
0957+561 will allow improved constraints on the lens mass
distribution and hence will improve the derived value of H$_0$. We
first present
improved optical positions and photometry for the known components of
this lens.  The optical
separation between the A and B quasar images agrees with VLBI data
at the 10~mas level, and the optical center of the primary lensing galaxy G1
coincides with the VLBI source G$^\prime$ to within 10~mas.
The best previous model for this lens (Grogin and Narayan 1996) is
excluded by these data and must be reevaluated.

Several new resolved features are found within 10\arcsec\ of G1, including
an apparent fold arc with two bright knots.  Several other small
galaxies are detected, including two which may be multiple images of
each other.  We present positions and crude photometry of these objects.
\end{abstract}
 
\section{Introduction}
 
Now that an accurate time delay between the two images of the
gravitationally lensed quasar Q0957+561 has been reliably determined
(Kundic \etal\markcite{K1} 1996), it should be possible to use this system to
measure H$_0$ to an accuracy of a few percent (Refsdal
\markcite{Re1} 1964).  This
measurement would be independent of the usual distance ladder and
insensitive to local motions.  The largest remaining uncertainty in
this method is our ignorance of the mass distribution in the lens.
This ignorance takes two forms: the first is a degeneracy in the lens
models under the addition of a smooth sheet of matter across the
central 10--20\arcsec\ of the lens.  Changing this smooth component in
a mass model leaves all image positions and magnifications unchanged,
yet alters the derived H$_0$.  Our first step in
reducing the uncertainty in the value of H$_0$ was to obtain a weak
lensing estimate of the smooth mass distribution contributed by the
small galaxy cluster at the primary lens redshift (Fischer
\etal\markcite{Fi1}
 1996).  Recent measurements of the velocity dispersion of G1
may also help reduce this degeneracy (Falco \etal\markcite{Fa2} 1997).
 
Our second form of ignorance is of the structure
of the mass distribution in the strong lensing region within 10\arcsec\
of G1.  Weak lensing statistical
mass reconstructions have insufficient resolution
to determine structure on these small scales.  
Existing models of the lens presume
some parametric form for the mass distribution of the primary lens
galaxy G1, and adjust the parameters until the quasar image positions
are reproduced ({\it e.g.\/} Falco, Gorenstein, \& Shapiro
\markcite{Fa1} 1991;
Bernstein, Tyson, \& Kochanek
\markcite{B1} 1994 [BTK94]; Grogin \& Narayan \markcite{Gr1} 1996).
Tighter and more numerous constraints on the optics of the lens
will lead to more detailed and precise models of the mass distribution,
and hence more accurate values of H$_0$.  The best existing constraints
are the highly accurate quasar core positions from VLBI observations
(Gorenstein \etal\markcite{Go2} 1988) and the further VLBI observations of the
structure of the 50~mas jets which extend from each quasar core
(Garrett \etal\markcite{Ga2} 1994).  
We have obtained deep WFPC2 observations of the 0957+561 system
to see if the visible-light positions of
the quasars match the VLBI positions, and to see which (if any) of
the reported weak radio sources corresponds to the center of G1.
Furthermore, nature has been kind, and the WFPC2 images reveal
other objects in the background
of G1 which are strongly lensed, allowing 
us to place new constraints on the mass distribution.
 
\section{Observations}  \label{observations}
 
Observations of the field around the double QSO 0957+561 were taken with the
WFPC2 on 19 Nov. 1995 and 26 Nov. 1995 with the QSO centered in WF3. Two
filters were employed, with $2 \times 80$s and $14 \times 2300$s in F555W and
$2 \times 160$s, $1\times 500$s and $2\times900$s in F814W. The images were
pipeline processed in the standard manner.  The observations were dithered by
an integer number of WF pixels with a maximum offset of $\pm 2\arcsec$ in
either dimension. The fourteen 2300s exposures were shifted, combined and sky
normalized using the Tukey biweight algorithm (Hoaglin, Mosteller, \& Tukey
\markcite{Ho1} 1983), which eliminates cosmic rays on the final combined
image. Figure~1a is the shallow 80~sec F555W image.  Figures~1b and 1c are the
deeper combined F555W image and are discussed below.

\end{multicols}
\begin{figure}
\epsscale{.85}
\plotone{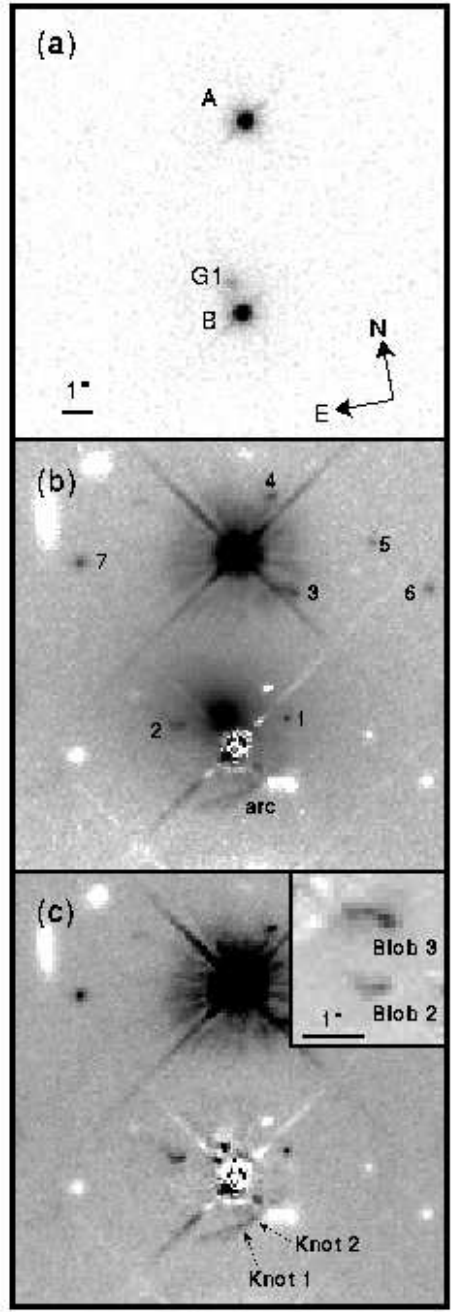}
\caption{Panel (a) is the 0957+561 lens system as seen in 80~second F555W WFPC2
images, with the A and B quasar images and primary lensing galaxy apparent.
Panel (b) shows a combine 39~ksec F555W image; a shifted version of the image
has been subtracted so that the A quasar acts as PSF template for B quasar.
The newly discovered arc and ``blobs'' are marked.  White objects are negative
``ghosts'' from the subtraction process.  Orientation and scale are same as
panel (a).  Panel (c) shows the same image after subtraction of the fitted
elliptical isophotes for G1.  The arc and its bright spots are more apparent;
other features in the vicinity of G1 are residuals from the B and G1
subtractions.  The inset shows zoomed images of Blobs 2 and 3, which may be
multiple images of the same background galaxy. \label{fig1}}
\end{figure}

\section{Astrometry} \label{astrometry}
 
With the resolution of the HST we should be able to determine whether
the optical quasar positions coincide with the cores of the VLBI radio
sources.  More importantly for lensing models, we wish to determine the
position of the center of galaxy G1, and see whether it coincides with
either the VLBI object G$^\prime$ (Gorenstein \etal\ 1988)
or the VLA object G (Roberts \etal\markcite{Ro1} 1985).
 
\subsection{Quasar Separation} \label{qsosep}
 
The centroids of the A and B quasars were determined on each of the four short
exposures, and the STSDAS task METRIC was used to remove the geometric
distortions on the WF3 chip and place the objects on the J2000 system.  The
vector from the B to the A quasar image has measured length
$r_{HST}=6\farcs169(3)$, at position angle ${\rm PA}_{HST}=-11\fdg4(1)$
(from N through E).  The uncertainties are dominated by calibration
factors:  the
pixel scale of WF3 is uncertain by $\approx3$ parts in $10^4$ (Cox
\markcite{Co2}
1996), and the PA is uncertain 
because the HST roll angle is calculated from the
positions of the two guide stars, which are known only to $\approx0.5\arcsec$
accuracy.
 
The most accurate measures of the (A-B) separation are from the VLBI
data of Gorenstein \etal\ (1988), who determine
$r_{VLBI}=6\farcs17499(2)$ and ${\rm PA}_{VLBI}=-11\fdg7029(2)$ in
epoch 1950.  After rotating the PA into J2000, the VLBI data yield
$r_{VLBI}=6\farcs175$ and ${\rm PA}_{VLBI}=-11\fdg453$, with
miniscule error compared to the HST data.  The HST and VLBI distances
from B to A agree to 6~mas ($2\sigma$), one part in $10^3$.  The PA
of the (A-B) vector agrees to within our 0\fdg1 knowledge of the HST
roll angle, which corresponds to a 10~mas uncertainty in the (A-B)
vector.  Because the agreement is good and the uncertainties in the
optical positions are dominated by calibration factors, we will
assume henceforth that the optical positions coincide with the VLBI
cores, and will use the VLBI (A-B) vector to calibrate the orientation and
scale of the HST image.  The uncertainties in the HST guide star
catalog preclude a useful comparison of the absolute positions of the
quasars in the optical vs.\ radio images.
 
\subsection{G1 Position} \label{G1}
 
The lensing galaxy G1 is visible on each short HST exposure, but the low S/N
and preponderance of cosmic rays make it more profitable to determine its
center from the combined deep HST image.  To successfully measure G1, it is
necessary to subtract the B quasar flux from the image.  We find that the A
quasar image makes a far better PSF template than can be produced by any other
method.  From the combined deep image we subtract a copy which is shifted so
that the A quasar overlapps B.  yielding the image shown in Figure~1b.
 
A robust estimate of the G1 center was made by centroiding and ellipse-fitting
over different apertures on the combined image.  We cannot measure the
positions of A and B on the combined deep image because they are saturated, so
we must transfer the short-exposure coordinate system (as determined by forcing
the (A-B) vector to agree with VLBI data) onto the combined image using the
8--10 objects which are visible on the short exposures yet unsaturated in the
deep images.  The errors in this transfer are comparable to the errors in the
G1 center.  The optical position of the center of G1 is (0\farcs1776(35) E,
1\farcs0186(35) N) (J1950) of quasar B.  This is fully consistent with the
ground-based position for G1 ($0.19\pm0.03\arcsec$ E, $1.00\pm0.03\arcsec$ N)
of B (Stockton \markcite{S1} 1980).  We list J1950 positions in this section
for easy comparison with previous works; the J2000 positions of the various
central objects are listed in Table~1.
 
The center of G1 is within 10~mas ($2.7\sigma$) of the VLBI source G$^\prime$
at (G$^\prime$-B)=(0\farcs181(1) E, 1\farcs029(1) N) reported by Gorenstein
\etal\markcite{Go1} (1983).  Independent EVN observations at 18~cm by Garrett
\markcite{Ga1} (1990) place G$^\prime$ at (0\farcs179(1) E, 1\farcs026(1) N)
relative to B (all are J1950).  Thus it seems that the G$^\prime$ radio source
is most likely at the center of G1; the 10~mas difference between G$^\prime$
and G1 centers, if real, is only $\approx40\,h^{-1}$~pc at the distance of G1.
If G$^\prime$ were the third image of the quasar core, we would expect it to be
demagnified relative to the A and B images by a factor $\sim
r_{G^\prime}/r_{E}$, where $r_{G^\prime}$ is 10~mas and $r_E$ is the Einstein
radius of the lens, roughly 3\arcsec.  Thus G$^\prime$, if a third image,
should have a flux $\sim0.3$\% that of the B image, whereas Garrett (1990)
measures a G$^{\prime}/$B flux ratio of 2.8\%.  Similar G$^\prime$/B flux
ratios were found by Gorenstein \etal\ (1983), suggesting that G$^\prime$ is
about an order of magnitude too bright to be the third image of the quasar.
 
The radio source G found in 6~cm VLA maps of the system (Roberts
\etal\ 1985) is significantly displaced from both our optical and the
VLBI sources: (G-B)=(0\farcs151(1) E, 1\farcs051(1) N).  Likewise
18~cm MERLIN observations by Garrett (1990) yield
(G-B)=(0\farcs148(8) E, 1\farcs093(8) N).  Both of these observations
find G many sigma to the North and West of G$^\prime$ and G1.  The
discrepancy is likely due to structure in the radio sources which is
not resolved by the VLA or MERLIN observations; the VLBI observations
clearly show such structure in the A and B images, and there may be
jets in G as well which are undetected by VLBI.  These issues are
discussed in detail in the radio observation papers.
 
Finally we note that the lens model for 0957+561 by Grogin \& Narayan
(1996) places the lens (G1) center at (0\farcs215 E, 1\farcs057 N) of
B, $15\sigma$ from the new optical position.  These authors note that
their model fits the quasar jet positions extremely poorly if the
galaxy is forced to reside near G$^\prime$, as indicated by the HST
images, so a reevaluation of the lens models is now required.
 
\section{Surface Photometry of G1} \label{surface}
The surface photometry task ISOPHOTE in the STSDAS package is used to fit
elliptical isophotes to G1.  Isophote fitting is done on the image shown in
Figure~1b, in which the A quasar has been shifted and subtracted from the B
quasar.  Note that this leaves a negative ``ghost'' quasar 6\arcsec\ south of
G1.  This ghost and the A quasar image limit the radius to which surface
photometry can be done.  During the ellipse fit we mask the areas covered by
the B quasar core and the newly discovered small objects described in
\S\ref{blobs}.  Figure~1c shows the WFPC2 image after subtraction of the fitted
model for G1.
 
The surface brightness, ellipticity, and PA of the G1 isophotes are plotted vs
radius in Figure~2.  We take the measured F555W surface brightnesses to be
equivalent to $V$ band, since Fukugita \etal\markcite{Fu1} (1995) calculate
F555W-$V$ for a $z=0.36$ elliptical to be $<0.03$~mag.  We have overplotted the
ground-based $R$-band surface photometry of G1 from BTK94, assuming the
predicted color $V-R=1.55$ that a present-day elliptical galaxy would have if
observed at $z=0.36$ (Coleman, Wu, and Weedman \markcite{Co1} 1980).  This
leaves the BTK94 data $\approx0.2$~mag fainter than the WFPC2 data in the
region of overlap.  This small discrepancy may be ascribed to passive evolution
in G1 and perhaps calibration errors.
 
The WFPC2 and KPNO datasets match smoothly in all measured parameters.
The HST surface photometry inside 2\arcsec\ radius continues the
trends seen farther out in the BTK94 profiles---a slight isophotal
twist and a tendency toward rounder isophotes in the center of G1
[wiggles in the ellipticity and PA profiles are probably due to
residual quasar flux].
The surface brightness profile is no longer well fit by
a single power law or by a deVaucoleurs profile; the latter is
too flat in the center of the galaxy.  Like most ellipticals,
G1 continues a power-law rise in surface brightness to the center
at the resolution limit of the HST (Gebhardt \etal\markcite{Ge1} 1996), which
means that the third quasar image is likely to be highly demagnified.

\end{multicols}
\begin{figure}
\plotone{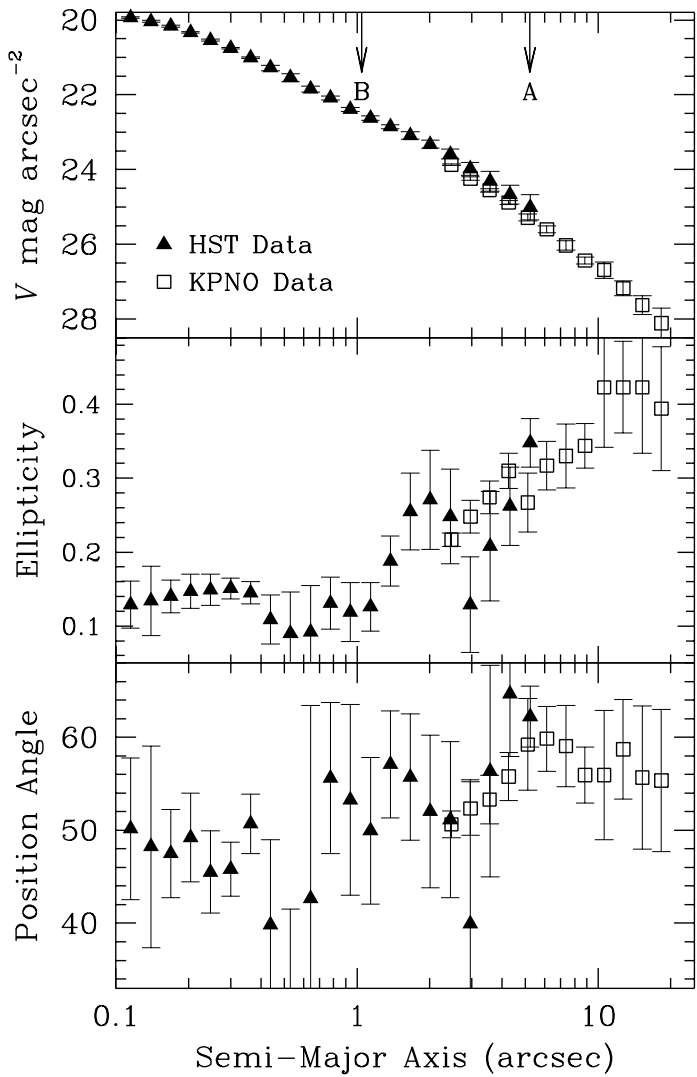}
\caption{Fitted elliptical isophotes for G1 are plotted vs the semi-major axis
of the isophotes.  Filled symbols are from fits to the WFPC2 image F555W image,
which should be very close to $V$-band.  Open symbols are from the ground-based
$R$-band image described in BTK94.  The upper panel shows the surface
brightness profile; the $R$ data are plotted assuming $V-R=1.55$ as expected
from nearby elliptical galaxies redshifted to $z=0.36$.  Middle and lower
panels show the ellipticity and PA of the fitted isophotes; most of the scatter
is due to residuals from the subtraction of quasar B, but in general we see
good agreement between the 2 data sets, a trend toward rounder isophotes in the
center, and a slight isophotal twist.  The ``A'' and ``B'' at the top axis mark
the radii of the A and B quasar images. \label{fig2}}
\end{figure}
 
\section{Newly Discovered Features} \label{blobs}
 
Several previously undiscovered objects are visible within
10\arcsec\ of G1 on the quasar-subtracted WFPC2 image (Figure~1b) and
the galaxy-subtracted image (Figure~1c).  At least some of these
are likely to be lensed background objects, which will place new
constraints on lens models for this system.  Table~2 lists positions,
total magnitudes, and peak surface brightnesses (at the WFPC2 resolution)
of the new features.
 
\subsection{Arc}
 
A previously undetected ``arc'' is visible just outside the
B QSO image. The arc is visible even before the subtraction of
the B QSO quasar flux, thus is not an artifact of the PSF subtraction.
There appears to be substructure in the arc, specifically two
bright spots (``knots'') separated by 0\farcs5.  Rough coordinates of
points along
the arc and of Knots 1 \& 2 are given in Table~2.  The arc is
of quite low surface brightness and the knots are faint ($V\approx26$)
and perhaps unresolved.
Because of the low S/N
of these features and the difficult QSO and galaxy subtraction, the
uncertainties on the knot positions and fluxes are only crudely estimated.
 
Unlike the putative
arc shown in BTK94, this candidate arc looks very much as one
would expect for a fold arc given the geometry of the QSO images.
Arcs are found astride critical lines, and for sources at redshifts
well behind the $z=0.36$ lens there must be a critical line
at a radius from G1 intermediate to the radii of the A and B quasar
images.  Indeed the new arc candidate is found at such an intermediate
radius.  The parts of a fold arc interior and exterior to the
critical line are mirror images of the same section of the source.
It seems likely that Knot 1 and Knot 2 are multiple images of the
nucleus of or a bright spot in a faint source galaxy.
The two knots have similar fluxes and similar peak surface brightnesses,
consistent with a fold arc scenario.  Higher-S/N images will show
more detail in the arc and knots, permitting a more definitive conclusion
as to the location of the critical line.
 
\subsection{Small Sources}
Seven other sources are visible in the vicinity of the A and B quasar
images.  These ``blobs'' are labelled in Figure~1b and their
positions, fluxes, and peak surface brightnesses  are
are listed in Table~2.
Uncertainties are difficult to quantify because
these are mostly extended objects and the sky subtraction is the
leading source of error.  We list all positional uncertainties as
0\farcs05, or one-half of a WFPC2 pixel.
 
Blob 1 is unresolved.  It is likely to be at the G1 redshift or
a foreground star, because if it were behind G1 it would almost
certainly be multiply imaged.  No other stellar objects are visible
in the strong lensing region.
 
Blob 2 is only 1\farcs4 from G1 and hence will be multiply imaged
if at redshift higher than G1.  Blob 2 is clearly resolved and hence
non-stellar.  Since the median redshift for $V=25$ galaxies is well
above $z=0.35$, it is very likely that a second image of Blob 2 is
present in the field.  A lens model which we fit to the quasar positions
suggests that Blob 3 has
the correct position and flux to be this second image.
The peak pixel of Blob 3 has $\sim2$ times higher flux than the peak
pixel in Blob 2, but this difference in apparent surface brightness
could be due to an unresolved bright spot in the source galaxy.
As can be seen in the inset to Figure~1c,
both galaxies are extended in the EW direction and are brighter
at their W ends; the lens model suggests that the counterimage of
Blob 2 should be flipped about a nearly horizontal axis, so their
crude morphologies are consistent with multiple imaging.  Higher-S/N
imaging will again allow a more definitive test of the multiple
imaging hypothesis.
 
The other four new objects are sufficiently distant from G1 that
they are probably not multiply imaged regardless of their redshifts.
 
\section{Conclusions}
 
The WFPC2 images provide a wealth of information useful in modelling
the strong-lensing effects in the 0957+561 system.  The optical
A and B quasar and G1 galaxy positions agree with the enormously precise VLBI
positions for A, B, and G$^\prime$
to within 10~mas.  This forces a revision of the
best published model of the 0957+561 lens (Grogin and Narayan 1996).
Fortunately the WFPC2 images provide new constraints to use in
modelling the system:  the arc constrains the location of the critical
line, especially if we can interpret the two bright knots as
multiple images; and the hypothesis that Blobs 2 and 3 are multiple
images of a common source adds another set of deflection and
magnification constraints to candidate models.  These new data add
at least 4 new constraint equations to models of this lens.
In a later publication
we will explore lens models which can satisfy these new constraints
as well as the position and flux constraints from previous radio
imaging.  The new STIS CCD imager on HST will be trained on this
system in Cycle 7 to provide higher S/N and better PSF sampling
on the newly discovered arc and background galaxies.  These images
should show sufficient detail in the arc and faint background
galaxies to definitively test the lensing hypothesis, to
determine the full relative magnification matrix for each lensed pair,
and perhaps to detect the fainter counter-images of the
arc, further increasing the constraints on the model.  This should
allow exquisite accuracy in the determination of the mass distribution
in this system, and improve the accuracy of the determination of $H_0$.

Support for this work was provided by NASA through grants \# HF-01069.01-94A
(PF) and \#GO-05979.0X-94A (GB, GR, JAT)
from the Space Telescope Science Institute, which is operated by the
Association of Universities for Research in Astronomy Inc., under NASA contract
NAS5-26555.
 
{}

\newpage
 
\begin{deluxetable}{ccllll}
\tablewidth{0pt}
\tablecaption{Positions for Central Object}
\tablehead{
\colhead{Object} & \colhead{Instrument} &
\colhead{RA\tablenotemark{1}} & \colhead{Dec\tablenotemark{1}} &
\colhead{Uncertainty} & \colhead{Reference}
}
\startdata
G1 & HST & 0\farcs1820 & 1\farcs0178 & 0\farcs0035 & This work \nl
G1 & CFHT & 0\farcs19 & 1\farcs00 & 0\farcs03 & Stockton 1980 \nl
G$^{\prime}$ & VLBI & 0\farcs185 & 1\farcs028 & 0\farcs001 &
 Gorenstein \etal\  1983 \nl
G$^{\prime}$ & EVN & 0\farcs183 & 1\farcs025 & 0\farcs001 &
Garrett 1990 \nl
G          & VLA & 0\farcs155 & 1\farcs050 & 0\farcs001 &
Roberts \etal\ 1985 \nl
\enddata
\tablenotetext{1}{Displacement from B center (J2000) listed.}
\end{deluxetable}

\begin{deluxetable}{cccccl}
\tablewidth{0pt}
\tablecaption{Faint Object Positions and Photometry}
\tablehead{
\colhead{Object} &
\colhead{RA\tablenotemark{1}} & \colhead{Dec\tablenotemark{1}} &
\colhead{Magnitude\tablenotemark{2}} & \colhead{Peak SB\tablenotemark{2}} &
\colhead{Remarks}
}
\startdata
Quasar A & $-1\farcs408$ & $+5\farcs034$ & \nodata & \nodata \nl
Quasar B & $-0\farcs182$ & $-1\farcs018$ & \nodata & \nodata \nl
Arc   &  $-1\farcs51$ & $-1\farcs81$ & \nodata & $\gtrsim 24.7$  &
Point along arc \nl
Arc   &  $-0\farcs93$ & $-2\farcs21$ & \nodata & $\gtrsim 24.7$  &
Point along arc \nl
Knot 2 &  $-0\farcs48$ & $-2\farcs43$ &
$25.8\pm0.5$ & $23.4\pm0.1$  & Position of peak flux \nl
Knot 1 &  $-0\farcs06$ & $-2\farcs55$ &
$26.1\pm0.5$ & $23.6\pm0.1$ & Position of peak flux\nl
Arc   &  $+0\farcs47$ & $-2\farcs66$ & \nodata & $\gtrsim 24.7$  &
Point along arc \nl
Arc   &  $+1\farcs09$ & $-2\farcs74$ & \nodata & $\gtrsim 24.7$  &
Point along arc \nl
Blob 1 & $-1\farcs95$ & $-0\farcs46$ & $25.6\pm0.1$ & $22.26\pm0.04$ &
Unresolved \nl
Blob 2 & $+1\farcs54$ & $-0\farcs05$ & $24.8\pm0.1$ & $22.95\pm0.07$ &
Possible counterimage of Blob 3 \nl
Blob 3 & $-2\farcs86$ & $+3\farcs47$ & $23.5\pm0.1$ & $22.32\pm0.04$ &
Possible counterimage of Blob 2 \nl
Blob 4 & $-2\farcs70$ & $+6\farcs54$ & $25.6\pm0.1$ & $22.64\pm0.05$  \nl
Blob 5 & $-5\farcs67$ & $+4\farcs58$ & $26.4\pm0.1$ & $23.6\pm0.1$  \nl
Blob 6 & $-7\farcs22$ & $+2\farcs77$ & $25.6\pm0.1$ & $23.2\pm0.1$  \nl
Blob 7 & $+3\farcs65$ & $+5\farcs66$ & $24.8\pm0.1$ & $22.48\pm0.04$  \nl
\enddata
\tablenotetext{1}{Displacement from G1 center (J2000) listed.  All positions
except Quasars uncertain by $\pm0\farcs05$.}
\tablenotetext{2}{F555W magnitudes are given;
$V$ mag should differ by $<0.05$ mag.}
\end{deluxetable}

\end{document}